\documentstyle [12pt,epsfig] {article}

\def\FigSize{0.7\textwidth}

\newcommand {\GLNC} {{GL(N,C)}}
\newcommand {\CP} {{\bf CP}}

\catcode`\@=11
\def\@citex[#1]#2{%
\if@filesw \immediate \write \@auxout {\string \citation {#2}}\fi
\@tempcntb\m@ne \let\@h@ld\relax \def\@citea{}%
\@cite{%
  \@for \@citeb:=#2\do {%
    \@ifundefined {b@\@citeb}%
      {\@h@ld\@citea\@tempcntb\m@ne{\bf ?}%
      \@warning {Citation `\@citeb ' on page \thepage \space
undefined}}%
      {\@tempcnta\@tempcntb \advance\@tempcnta\@ne%
      \@tempcntb\number\csname b@\@citeb \endcsname \relax%
      \ifnum\@tempcnta=\@tempcntb 
	\ifx\@h@ld\relax%
	  \edef \@h@ld{\@citea\csname b@\@citeb\endcsname}%
	\else%
	  \edef\@h@ld{\ifmmode{-}\else--\fi\csname
b@\@citeb\endcsname}%
	\fi%
      \else
	\@h@ld\@citea\csname b@\@citeb \endcsname%
	\let\@h@ld\relax%
      \fi}%
    \def\@citea{,\penalty\@highpenalty\,}%
  }\@h@ld
}{#1}}

\def\@citeb#1#2{{[#1]\if@tempswa , #2\fi}}
\def\@citeu#1#2{{$^{#1}$\if@tempswa , #2\fi }}
\def\@citep#1#2{{#1\if@tempswa , #2\fi}}

\def\bcites{         
	\catcode`\@=11
	\let\@cite=\@citeb
	\catcode`\@=12
}

\def\upcites{         
	\catcode`\@=11
	\let\@cite=\@citeu
	\catcode`\@=12
}

\def\plaincites{      
	\catcode`\@=11
	\let\@cite=\@citep
	\catcode`\@=12
}

\newcount\hour
\newcount\minute
\newtoks\amorpm
\hour=\time\divide\hour by 60
\minute=\time{\multiply\hour by 60 \global\advance\minute by-\hour}
\edef\standardtime{{\ifnum\hour<12 \global\amorpm={am}%
	\else\global\amorpm={pm}\advance\hour by-12 \fi
	\ifnum\hour=0 \hour=12 \fi
	\number\hour:\ifnum\minute<10
0\fi\number\minute\the\amorpm}}
\edef\militarytime{\number\hour:\ifnum\minute<10
0\fi\number\minute}

\def\draftlabel#1{{\@bsphack\if@filesw {\let\thepage\relax
   \xdef\@gtempa{\write\@auxout{\string
      \newlabel{#1}{{\@currentlabel}{\thepage}}}}}\@gtempa
   \if@nobreak \ifvmode\nobreak\fi\fi\fi\@esphack}
	\gdef\@eqnlabel{#1}}
\def\@eqnlabel{}
\def\@vacuum{}
\def\marginnote#1{}
\def\draftmarginnote#1{\marginpar{\raggedright\scriptsize\tt#1}}
\def\draft{
	\pagestyle{plain}
	\overfullrule=2pt
	\oddsidemargin -.5truein
	\def\@oddhead{\sl \phantom{\today\quad\militarytime} \hfil
	\smash{\Large\sl DRAFT} \hfil \today\quad\militarytime}
	\let\@evenhead\@oddhead
	\let\label=\draftlabel
	\let\marginnote=\draftmarginnote
	\def\ps@empty{\let\@mkboth\@gobbletwo
	\def\@oddfoot{\hfil \smash{\Large\sl DRAFT} \hfil}
	\let\@evenfoot\@oddhead}

\def\@eqnnum{(\theequation)\rlap{\kern\marginparsep\tt\@eqnlabel}%
	\global\let\@eqnlabel\@vacuum}  }

\def\blackfonts{
	\font\blackboard=msbm10 scaled\magstep1
	\font\blackboards=msbm8
	\font\blackboardss=msbm6
}

\def\nblack{            
	\def\ZZ{{Z \n{10} Z}}
	\def\NN{{N \n{14} N}}
	\def\CC{{C \n{11} C}}
	\def\RR{{R \n{11} R}}
	\def\QQ{{Q \n{12} Q}}
	\def\PP{{P \n{11} P}}
}

\def\prep{         
	\catcode`\@=11
	\input art10.sty
	\catcode`\@=12
	
	\let\small\null
	\def\blackfonts{
		\font\blackboard=msbm10
		\font\blackboards=msbm7
		\font\blackboardss=msbm5
	}
	\let\sl\it
	\twocolumn
	\sloppy
	\voffset=-2.54truecm
	\hoffset=-2.54truecm
	\flushbottom
	\parindent 1em
	\leftmargini 2em
	\leftmarginv .5em
	\leftmarginvi .5em
	\marginparwidth 48pt
	\marginparsep 10pt
	\setlength{\columnsep}{2truecm}
	\setlength{\textwidth}{25.4truecm}
	\setlength{\textheight}{17truecm}
	\baselineskip=16pt
	\oddsidemargin .18truein
	\evensidemargin .17truein
}

\def\eqalign#1{\null\,\vcenter{\openup\jot\m@th
  \ialign{\strut\hfil$\displaystyle{##}$&$\displaystyle{{}##}$\hfil
      \crcr#1\crcr}}\,}
\def\eqalignno#1{\displ@y \tabskip\centering
  \halign
to\displaywidth{\hfil$\@lign\displaystyle{##}$\tabskip\z@skip
    &$\@lign\displaystyle{{}##}$\hfil\tabskip\centering
    &\llap{$\@lign##$}\tabskip\z@skip\crcr
    #1\crcr}}

\def\section{\@startsection {section}{1}{\z@}{3.ex plus 1ex minus
 .2ex}{2.ex plus .2ex}{\large\bf}}
\def\subsection{\@startsection{subsection}{2}{\z@}{2.75ex plus 1ex
minus
 .2ex}{1.5ex plus .2ex}{\bf}}

\def\appendix{{\newpage\section*{Appendix}}\let\appendix\section%
	{\setcounter{section}{0}
	\gdef\thesection{\Alph{section}}}\section}

\def\abstract{\if@twocolumn
\section*{Abstract}
\else 
\begin{center}
{\bf Abstract\vspace{-.5em}\vspace{0pt}}
\end{center}
\quotation
\fi}

\catcode`\@=12

\def\ibid{{\it ibid.\/}}

\newcommand{\beq}{\begin{equation}}
\newcommand{\eeq}{\end{equation}}
\newcommand{\beqa}{\begin{eqnarray}}
\newcommand{\eeqa}{\end{eqnarray}}

\newcommand{\R}{{R}}
\newcommand{\C}{{C}}

\newcommand{\vk}{\vec{k}}

\def\noj#1,#2,{{\bf #1} (19#2)\ }
\def\jou#1,#2,#3,{{\sl #1\/ }{\bf #2} (19#3)\ }
\def\ann#1,#2,{{\sl Ann.\ Physics\/ }{\bf #1} (19#2)\ }
\def\cmp#1,#2,{{\sl Comm.\ Math.\ Phys.\/ }{\bf #1} (19#2)\ }
\def\ma#1,#2,{{\sl Math.\ Ann.\/ }{\bf #1} (19#2)\ }
\def\jd#1,#2,{{\sl J.\ Diff.\ Geom.\/ }{\bf #1} (19#2)\ }
\def\invm#1,#2,{{\sl Invent.\ Math.\/ }{\bf #1} (19#2)\ }
\def\cq#1,#2,{{\sl Class.\ Quantum Grav.\/ }{\bf #1} (19#2)\ }
\def\cqg#1,#2,{{\sl Class.\ Quantum Grav.\/ }{\bf #1} (19#2)\ }
\def\ijmp#1,#2,{{\sl Int.\ J.\ Mod.\ Phys.\/ }{\bf A#1} (19#2)\ }
\def\jmphy#1,#2,{{\sl J.\ Geom.\ Phys.\/ }{\bf #1} (19#2)\ }
\def\jams#1,#2,{{\sl J.\ Amer.\ Math.\ Soc.\/ }{\bf #1} (19#2)\ }
\def\grg#1,#2,{{\sl Gen.\ Rel.\ Grav.\/ }{\bf #1} (19#2)\ }
\def\mpl#1,#2,{{\sl Mod.\ Phys.\ Lett.\/ }{\bf A#1} (19#2)\ }
\def\nc#1,#2,{{\sl Nuovo Cim.\/ }{\bf #1} (19#2)\ }
\def\np#1,#2,{{\sl Nucl.\ Phys.\/ }{\bf B#1} (19#2)\ }
\def\pl#1,#2,{{\sl Phys.\ Lett.\/ }{\bf #1B} (19#2)\ }
\def\pla#1,#2,{{\sl Phys.\ Lett.\/ }{\bf #1A} (19#2)\ }
\def\pr#1,#2,{{\sl Phys.\ Rev.\/ }{\bf #1} (19#2)\ }
\def\prd#1,#2,{{\sl Phys.\ Rev.\/ }{\bf D#1} (19#2)\ }
\def\prl#1,#2,{{\sl Phys.\ Rev.\ Lett.\/ }{\bf #1} (19#2)\ }
\def\prp#1,#2,{{\sl Phys.\ Rept.\/ }{\bf #1C} (19#2)\ }
\def\ptp#1,#2,{{\sl Prog.\ Theor.\ Phys.\/ }{\bf #1} (19#2)\ }
\def\ptpsup#1,#2,{{\sl Prog.\ Theor.\ Phys.\/ Suppl.\/ }{\bf #1}
(19#2)\ }
\def\rmp#1,#2,{{\sl Rev.\ Mod.\ Phys.\/ }{\bf #1} (19#2)\ }
\def\yadfiz#1,#2,#3[#4,#5]{{\sl Yad.\ Fiz.\/ }{\bf #1} (19#2) #3%
\ [{\sl Sov.\ J.\ Nucl.\ Phys.\/ }{\bf #4} (19#2) #5]}
\def\zh#1,#2,#3[#4,#5]{{\sl Zh..\ Exp.\ Theor.\ Fiz.\/ }{\bf #1}
(19#2) #3%
\ [{\sl Sov.\ Phys.\ JETP\/ }{\bf #4} (19#2) #5]}

\def\beq{\begin{equation}}
\def\eeq{\end{equation}}
\def\beqar{\begin{eqnarray}}
\def\eeqar{\end{eqnarray}}
\def\non{\nonumber}

\def\nfrac#1#2{{\displaystyle{\vphantom1\smash{\lower.5ex\hbox{\sma
ll$#1$}}%
	\over\vphantom1\smash{\raise.25ex\hbox{\small$#2$}}}}}

\def\p#1{\mskip#1mu}
\def\n#1{\mskip-#1mu}
\def\stop{\p6.}
\def\comma{\p6,}

\def\to{\rightarrow}

\def\lae{\mathrel{\mathop{\smash{\lower .5 ex
\hbox{$\stackrel<\sim$}}}}}
\def\lae{\mathrel{\mathop{\smash{\lower .5 ex
\hbox{$\stackrel>\sim$}}}}}

\def\l:{\mathopen{:}\,}
\def\r:{\,\mathclose{:}}

\def\CM{{\cal M}}

\catcode`\@=11
\def\theequation{\arabic{equation}}
\catcode`\@=12

\nblack
\bcites

\nblack

\catcode`\@=11
\def\theequation{\thesection.\arabic{equation}}
\@addtoreset{equation}{section}
\@addtoreset{footnote}{section}
\@addtoreset{footnote}{subsection}
\catcode`\@=12

\newcommand{\beqn}{\begin{equation}}
\newcommand{\eeqn}{\end{equation}}
\newcommand{\beqnarray}{\begin{eqnarray}}
\newcommand{\eeqnarray}{\end{eqnarray}}

\newcommand{\tQ}{{\tilde{Q}}}

\newcommand{\cM}{{\cal M}}

\newcommand {\nono} {\nonumber \\}

\newcommand {\bear} [1] {\begin {array} {#1}}
\newcommand {\ear} {\end {array}}

\newcommand {\eqr} [1]
	{(eq. \ref {eq:#1})}

\newcommand {\beqarn} {\begin{eqnarray*}}
\newcommand {\eeqarn} {\end{eqnarray*}}

\newcommand	{\abs}	[1] {{\left| #1 \right|}}

\newcommand {\mrm} [1] {\mathrm {#1}}

\newcommand {\comm} [2] {\mbox {$\left[ #1, #2 \right]$}}

\newcommand {\tr} {{\mbox {$\mrm{tr}$}}}

\newcommand {\myref} [1]	%
	{%
	\begin{thebibliography} {99}	%
			{#1}	%
	\end {thebibliography}}

\def\emph#1{{\em #1}}
\def\mathbf#1{{\bf #1}}
\def\mathrm#1{{\rm #1}}
\def\mathit#1{{\it #1}}

\newcommand {\pa}	{{\partial}}

\catcode`\@=11
\@addtoreset{footnote}{section}
\catcode`\@=12

\newcounter	{exinsert}	[subsection]

\renewcommand {\theexinsert}	%
{	\thesubsection.\arabic{equation}}

\renewcommand {\theequation} {\thesection.\arabic{equation}}

\catcode`\@=11
\@addtoreset{equation}{section}
\catcode`\@=12

\newenvironment {exinsert} [1]	%
{	
	\begin {quotation}	%
	\refstepcounter {equation}	%
	{\bf {#1}} \theequation. \,\,	%
}
{	
	\end {quotation}
}

\newcommand {\bprop}
{\begin {exinsert} {Proposition}
}
\newcommand {\eprop} {\end {exinsert} }

\newcommand {\bexe} {\begin {exinsert} {Exercise}}
\newcommand {\eexe} {\end {exinsert} }

\newcommand {\bexa} {\begin {exinsert} {Example}}
\newcommand {\eexa} {\end {exinsert} }

\newcommand {\literal} [1] {{ \mbox {$\mrm {#1}$}}}

\newcommand {\Rank} [0] {{\literal {Rank}}}

\newcommand {\grad} {\nabla}

\newcommand {\ncmd} [2] {\newcommand {#1} {#2}}

\ncmd {\vgrad} {{\vec \grad}}

\ncmd {\Mv} {{\cal M}_V}
\ncmd {\Mh} {{\cal M}_H}

\ncmd {\adj} {\literal {adj}}	
\ncmd {\Mp} {M_\literal {planck}}	
\def\FI{Fayet-Iliopoulos }

\ncmd {\x} {$\times$}

\newcommand {\myfigure} [4]
  {\begin{figure} [htb]
   \begin{center}
	\epsfig{figure=#2, width=#3}
   \end{center}
   \caption {#4}	\label {fig:#1}
  \end{figure}}

\newcommand {\putfig} [3] {\myfigure {#1} {#1.eps} {#2} {#3}}

\newcommand {\figr} [1]	{figure \ref {fig:#1}}

\newcommand {\btab} [3]
  {\begin{table} [htb]
   \begin{center}
   \caption {#2}	\label {tab:#1}
   \begin {tabular} {#3}
   \hline}

\newcommand {\etab}	{   \hline
   \end {tabular}
   \end{center}
  \end{table}}

\newcommand {\tabr} [1]	{table \ref {tab:#1}}

\ncmd {\bR} {{R}}
\ncmd {\bC} {{C}}

\ncmd {\CQ} {\cal Q}

\def\jou#1,#2,#3,{{\sl #1\/ }{\bf #2} (19#3)\ }
\def\ibid#1,#2,{{\sl ibid.\/ }{\bf #1} (19#2)\ }
\def\am#1,#2,{{\sl Acta. Math.\/ } {\bf #1} (19#2)\ }
\def\annp#1,#2,{{\sl Ann.\ Physics\/ }{\bf #1} (19#2)\ }
\def\cmp#1,#2,{{\sl Comm.\ Math.\ Phys.\/ }{\bf #1} (19#2)\ }
\def\cqg#1,#2,{{\sl Class.\ Quantum Grav.\/ }{\bf #1} (19#2)\ }
\def\dm#1,#2,{{\sl Duke\ Math.\ J.\/ }{\bf #1} (19#2)\ }
\def\ijmpa#1,#2,{{\sl Int.\ J.\ Mod.\ Phys.\/ }{\bf A#1} (19#2)\ }
\def\invm#1,#2,{{\sl Invent.\ Math.\/ }{\bf #1} (19#2)\ }
\def\jhep#1,#2,{{\sl JHEP\/ }{\bf #1} (19#2)\ }
\def\jmp#1,#2,{{\sl J.\ Geom.\ Phys.\/ }{\bf #1} (19#2)\ }
\def\jams#1,#2,{{\sl J.\ Diff.\ Geom.\/ }{\bf #1} (19#2)\ }
\def\jdg#1,#2,{{\sl J.\ Amer.\ Math.\ Soc.\/ }{\bf #1} (19#2)\ }
\def\jpa#1,#2,{{\sl J.\ Phys.\ A.\/ }{\bf #1} (19#2)\ }
\def\grg#1,#2,{{\sl Gen.\ Rel.\ Grav.\/ }{\bf #1} (19#2)\ }
\def\mpla#1,#2,{{\sl Mod.\ Phys.\ Lett.\/ }{\bf A#1} (19#2)\ }
\def\ma#1,#2,{{\sl Math.\ Ann.\/ } {\bf #1} (19#2)\ }
\def\nc#1,#2,{{\sl Nuovo\ Cim.\/ }{\bf #1} (19#2)\ }
\def\npb#1,#2,{{\sl Nucl.\ Phys.\/ }{\bf B#1} (19#2)\ }
\def\plb#1,#2,{{\sl Phys.\ Lett.\/ }{\bf #1B} (19#2)\ }
\def\pla#1,#2,{{\sl Phys.\ Lett.\/ }{\bf #1A} (19#2)\ }
\def\pnas#1,#2,{{\sl Proc.Nat.Acad.Sci.\/ }{\bf #1} (19#2)\ }
\def\prev#1,#2,{{\sl Phys.\ Rev.\/ }{\bf #1} (19#2)\ }
\def\prd#1,#2,{{\sl Phys.\ Rev.\/ }{\bf D#1} (19#2)\ }
\def\prl#1,#2,{{\sl Phys.\ Rev.\ Lett.\/ }{\bf #1} (19#2)\ }
\def\prpt#1,#2,{{\sl Phys.\ Rept.\/ }{\bf #1C} (19#2)\ }
\def\ptp#1,#2,{{\sl Prog.\ Theor.\ Phys.\/ }{\bf #1} (19#2)\ }
\def\ptpsup#1,#2,{{\sl Prog.\ Theor.\ Phys.\/ Suppl.\/ }{\bf #1} (19#2)\ }
\def\rmp#1,#2,{{\sl Rev.\ Mod.\ Phys.\/ }{\bf #1} (19#2)\ }
\def\sm#1,#2,{{\sl Selec. Math.\/ }{\bf #1} (19#2)\ }
\def\yadfiz#1,#2,#3[#4,#5]{{\sl Yad.\ Fiz.\/ }{\bf #1} (19#2) #3%
\ [{\sl Sov.\ J.\ Nucl.\ Phys.\/ }{\bf #4} (19#2) #5]}
\def\zh#1,#2,#3[#4,#5]{{\sl Zh.\ Exp.\ Theor.\ Fiz.\/ }{\bf #1} (19#2) #3%
\ [{\sl Sov.\ Phys.\ JETP\/ }{\bf #4} (19#2) #5]}

\begin {document}


\begin{titlepage}

\begin{center}
\today
\hfill LBNL-41466, UCB-PTH-98/13\\
\hfill                 hep-th/9803050

\vskip 1.5 cm
{\large \bf Matrix Description of Intersecting M5 Branes}
\vskip 1 cm
{Shamit Kachru\footnote{Present address: Institute for Theoretical
Physics, University of California, Santa Barbara, CA 93106}, Yaron Oz and
Zheng Yin}\\
\vskip 0.5cm
{\sl Department of Physics,
University of California at Berkeley\\
366 Le\thinspace Conte Hall, Berkeley, CA 94720-7300, U.S.A.\\
and\\
Theoretical Physics Group, Mail Stop 50A--5101\\
Ernest Orlando Lawrence Berkeley National Laboratory, \\
Berkeley, CA 94720, U.S.A.\\}

\end{center}

\vskip 0.5 cm
\begin{abstract}
Novel 3+1 dimensional ${\cal N}=2$ superconformal field theories
(with tensionless BPS string solitons) are believed to arise
when two sets of M5 branes intersect over a 3+1 dimensional
hyperplane.  We derive a DLCQ description of these theories
as supersymmetric quantum mechanics on the Higgs branch of
suitable 4d ${\cal N}=1$ supersymmetric gauge theories.
Our formulation allows us to determine the
scaling dimensions of certain chiral primary operators
in the conformal field theories.
We also discuss general criteria for quantum
mechanical DLCQ descriptions of supersymmetric field theories
(and the resulting multiplicities and scaling dimensions of chiral primary
operators).

\end{abstract}
\end{titlepage}

\section{Introduction}

Many new nontrivial RG fixed points of supersymmetric field 
theories in various dimensions have
been discovered in
recent years.  
A host of novel fixed points in 3,4,5, and 6 dimensions
were discovered using string theory arguments.
For many of these theories, there is no known ultraviolet
Lagrangian which flows to them in the infrared.  Therefore,
it is of interest to find other ways of defining them, that
do not involve all of the degrees of freedom of string or M
theory.

For some of the simplest novel fixed points in six dimensions,
with $(2,0)$ and $(1,0)$ supersymmetry, such an alternative
definition has been proposed in \cite{abkss,witten,lowe,abks}.
In analogy with the matrix model for M theory \cite{bfss,susskind},
it was proposed that the discrete light-cone quantization (DLCQ)
of these 6d field theories can be formulated in terms of a suitable
supersymmetric quantum mechanics.  A similar description of 4d
${\cal N}=4$ Super Yang-Mills was discussed in \cite{orisav}.  In this
paper, we initiate the study of matrix descriptions for 4d theories
with 8 supercharges, by providing an analogous DLCQ description of
a class of ${\cal N}=2$ superconformal fixed points.
These 3+1 dimensional fixed points govern the physics on the
intersection of $K$ M5 branes intersecting $K'$ M5 branes
along a 3+1 dimensional hyperplane.  The two sets of M5 branes
can be connected by membranes which give rise to BPS
saturated tensionless strings 
on the intersection.  Such intersections were discussed, for instance, 
in \cite{hankleb}.

In the next section, we present the brane configuration which 
gives rise to the ${\cal N}=2$ fixed points and specify the
decoupling limit (in which the physics on the brane intersection
should be expected to decouple from gravity). 
We  construct a  Matrix description of this system, 
by studying the quantum mechanics
of $N$ D0 branes in the background of intersecting D4 branes. 
The zero brane quantum mechanics reduces to a
sigma model on the Higgs branch of a 4d ${\cal N}=1$ supersymmetric
gauge theory. 
In \S3 we study the structure of the Higgs 
branch for general $K,K'$ and provide arguments for the decoupling. 
We find that the quantum mechanics has  
a branch 
localized on the intersection which decouples from
the ``bulk'' in the limit of \S2.
In \S4 we analyze     
quantum mechanical states that correspond to chiral primary operators of the 4d 
superconformal theory.
These states come from compact cohomology 
representatives localized at the origin of the Higgs branch, as in \cite{abs}.
By computing this cohomology, we are able to provide the multiplicities
and scaling dimensions of certain chiral primary operators in the conformal
field theory.
In \S5 we discuss general 
criteria for a Matrix description of supersymmetric field theories, 
and make some general remarks about
the resulting multiplicities and scaling dimensions of   
chiral primary operators.
In \S6, we summarize the main points and discuss relations with other
recent work on the DLCQ description of field theories.


\ncmd {\bI} {\textbf {I}}
\ncmd {\id} {\bC}

\ncmd {\Sym} {{\literal {Sym}}}

\section {The Target and Its Probe}

\subsection {The Target: A Theory of Tensionless Strings}

	We start in M theory with a 
number of M5 branes, whose
worldvolume configurations
can be divided into two types that we label as
M5 and M5' in \tabr {brane-config-M}.  There are
$K$ and $K'$ of them respectively.
\btab {brane-config-M} {Configurations of the branes in M theory}
	{ l  c  c  c  c  c  c  c  c  c  c  c }
 & 0 & 1 & 2 & 3 & 4 & 5 & 6 & 7 & 8 & 9 & 10\\ \hline
M5 & \x & \x & \x & \x & \x & & & & & & \x \\ \hline
M5' & \x & \x & \x & & & \x & \x & & & & \x \\ \hline
\etab
Such a configuration can preserve up to 8 supercharges, corresponding
to ${\cal N}$=2 in 4d.  Of the original $Spin(1,10)$ Lorentz symmetry, only
the (1+3)d Lorentz group, $SL(2,\bC)$, of the 0, 1, 2, and 10th directions
and the Euclidean
rotation group $SU(2)_{789} \times U(1)_{34} \times U(1)_{56}$
remain manifest.  From the usual rule for branes ending on branes 
\cite {strominger-open-p,townsend-brane-surgery}, we know that
there can be open membranes ending on and stretched in between
the two types of M5-branes.  They look
like strings in the (1+3)d common directions and have tension
$M_{pl}^3 l$.  We are interested in the limit
\beq	\label {eq:decoupling-limit}
	M_{pl} \to \infty
\eeq
and
\beq	\label {eq:tensionless-limit}
	M_{pl}^3 l = {\rm fixed},
\eeq
where $l$ is the distance between the two sets of M5 branes
in the 7-8-9 directions.
In this double scaling limit, the bulk gravity decouples from the M5 branes
while the tension of the BPS strings mentioned above remains constant.
Our particular interest is in the limit when the two sets of M5 branes
coincide, and the BPS strings become tensionless.
One
might expect a theory with tensionless string solitons to be
nontrivial.  It is known that the decoupled theory on two parallel
and coincident M5-branes is interacting \cite {abkss,witten},  and we expect
that the configuration in \tabr {brane-config-M} also yields
interacting fixed points.  What is {a priori} not
obvious is whether the tensionless strings in this case are a feature of
an intrinsic 3+1 dimensional theory, localized at the
intersection and decoupled from the ``bulk'' of the two types of M5 branes.
In latter sections we will present evidence in support of this.

	We can compactify
the 10th direction on a circle and go to type IIA string theory.
The resulting configuration is that of two sets of D4-branes, as summarized in
\tabr {brane-config-IIA}.
\btab {brane-config-IIA} {Configurations of the branes in IIA}
	{ l  c  c  c  c  c  c  c  c  c  c }
 & 0 & 1 & 2 & 3 & 4 & 5 & 6 & 7 & 8 & 9 \\ \hline
D4 & \x & \x & \x & \x & \x & & & & & \\ \hline
D4' & \x & \x & \x & & & \x & \x & & & \\ \hline
\etab
The strings from open membranes can be either wrapped around
the 10th direction or transverse to it
in the M theory picture.  In the IIA picture these two kinds of
configurations give rise to
particles from open strings and strings from open D2-branes.
The particles make up $K \times K'$ hypermultiplets.
In the limit \eqr {tensionless-limit}, they are massless.  We shall
label the scalars in them and their VEVs as $\theta_{a}^{a \prime}$ and
$\tilde \theta^{a}_{a \prime}$, $K \times K'$ and $K' \times K$ matrices
respectively.

\subsection {The Probe}

	As usual it is very difficult to analyze this interacting
system using conventional field theory techniques.  To this end
we take the DLCQ approach of \cite {abkss,witten,abs}.
After the procedures outlined in \cite {seiberg-why-correct,sen}, the
physics of
the N momentum sector is described by N D0-branes
probing the configuration of \tabr {brane-config-IIA}.
The DLCQ procedure breaks the $SL(2, \bC)$ Lorentz group of
\tabr {brane-config-M} down to $U(1)_{12}$.  This combined system now
has 4 supercharges, equivalent to ${\cal N}$=1 supersymmetry in 4d reduced
to quantum mechanics.  The transformation properties of the holomorphic
supercharges $\CQ$ and the superpotential are given in \tabr {supercharge}.
To find the lightest fields of the
quantum mechanics and their interactions, we consider first the D0
brane probes alone and then introduce
the two types of D4-branes.
\btab {supercharge}
{Transformation property of the supercharges and superpotential}
{ l  c  c  c  c }
 & $U(1)_{12}$ & $U(1)_{34}$ & $U(1)_{56}$ & $SU(2)_{789}$  \\ \hline
$\CQ$ & +1 & +1 & +1 & (2) \\ \hline
$W$ & +2 & +2 & +2 & (1) \\ \hline
\etab

	From the D0-branes themselves, we have the content of an 
${\cal N}$=4 D=4 vector multiplet dimensionally reduced to quantum mechanics.
This breaks down to one ${\cal N}$=1 D=4 vector multiplet and 3 chiral
multiplets in the adjoint.  In total there are 9 scalars,
$\Phi^i$, $i = 1\ldots9$, parameterizing
the transverse fluctuation of the D0-brane.
The scalars in the vector multiplet in the quantum mechanics
are $\Phi^7, \Phi^8, \Phi^9$ (in the language of the dimensional
reduction, they come from Wilson lines of the gauge field around the
$T^3$).
The other scalars come from the
dimensional reduction of the 3 chiral multiplets.
We write their holomorphic
combinations as $\Phi_{12}$, $\Phi_{34}$, and $\Phi_{56}$
respectively.  There is a superpotential
\[	W_{PP} = tr \Phi_{12} \comm {\Phi_{34}} {\Phi_{56}}.	\]

	The subsystem consisting of the probes and the unprimed
D4-branes supports 8 supercharges, dimensionally
reduced  from ${\cal N}$=2 in 4d.  In addition to the fields described
above, we also have $K$ hypermultiplets in the fundamental
of $U(N)$, coming from open strings starting on the D0 branes and ending
the D4 branes.  They decomposes
under the supersymmetry
in \tabr {supercharge} into
chiral multiplets $Q$ and $\tilde Q$ with $a = 1 \ldots K$.
This system has a superpotential
\beq
	W_{PA} = \sum_a Q \Phi_{56} \tilde Q.
\eeq
Similarly for the subsystem consisting of the probes and the primed
D4-branes, we obtain chiral multiplets $Q'$ and $\tilde Q'$
with $a'= 1 \ldots K'$ and a superpotential
\beq
	W_{PB} = \sum_{a \prime} Q' \Phi_{34} \tilde Q'.
\eeq

To summarize, the fields, parameters and R
charges in the quantum mechanical theory are as follows:

\btab {field} {Fields of the Quantum Mechanics}
{ l  c  c  c  c }
 & $U(1)_{12}$ & $U(1)_{34}$ & $U(1)_{56}$ & $SU(2)_{789}$  \\ \hline
$\Phi_{12}$ & +2 & 0 & 0 & (1) \\ \hline
$\Phi_{34}$ & 0 & +2 & 0 & (1) \\ \hline
$\Phi_{56}$ & 0 & 0 & +2 & (1) \\ \hline
$Q, \tilde Q$ & $+1$ & +1 & 0 & (1) \\ \hline
$Q', \tilde Q'$ & +1 & 0 & +1 & (1) \\ \hline
$\Phi_{789}$ & 0 & 0 & 0 & (3) \\ \hline
\etab

In keeping with the usual matrix approach, VEVs of fields in the target
(D4-brane)
theory
becomes parameters in the probe quantum mechanics.
The obvious ones are the diagonal
VEVs for the scalars in the adjoint of
$U(K)$ and $U(K')$ from the two sets of D4-branes, which we will call
$X$ and $X'$ respectively.  They are mass parameters
and can be shifted to zero for the configuration satisfying
\eqr {tensionless-limit}.  Then there are the VEVs for the hypermultiplets
$\theta$ and $\tilde \theta$, as well as backgrounds for antisymmetric
3-form tensor field strengths $H$ and $H'$ on the two types of M5-branes.
The latter map to \FI parameters in the quantum mechanics, as
in \cite {abs}.  For the model
under consideration here, the FI parameters split into real parameters
$\zeta_\R$ and $\zeta'_\R$ as well as complex parameters
$\zeta_\C$ and $\zeta'_\C$.
In addition to these parameters, there is also the coupling constant $g$ for
the D0-brane gauge theory.  It is given by
\beq
	g^2 = (R M_{pl}^2)^3
\eeq
where $R$ is the radius of the DLCQ circle.  $g$ has mass dimension $\frac
3 2$.
The limit \eqr {decoupling-limit} implies that $g \to \infty$ and therefore
we are interested in the infrared (i.e. large time) limit of the probe
quantum mechanics.  Unlike the gauge coupling for ${\cal N}$=1 supersymmetric
gauge theories in 4d, the gauge coupling $g$ is not 
part of a background field that is
charged under the $U(1)$ global symmetries.
The transformation properties of all these
parameters are given in \tabr {parameter}.
\btab {parameter} {Parameters of the Quantum Mechanics}
{ l  c  c  c  c }
 & $U(1)_{12}$ & $U(1)_{34}$ & $U(1)_{56}$ & $SU(2)_{789}$  \\ \hline
$\zeta_{\bR}$ & 0 & 0 & 0 & (1) \\ \hline
$\zeta_A$ & +2 & +2 & 0 & (1) \\ \hline
$\zeta_B$ & +2 & 0 & +2 & (1) \\ \hline
$\theta, \tilde \theta$ & 0 & +1 & +1 & (1) \\ \hline
\etab
By the usual supersymmetric nonrenormalization theorem, $g$ can only
appear in the superpotential through nonperturbative effects.
Since $g$ carries no charges under any global symmetry, there is no
other restriction on it.   From this table and from the analysis
in \cite {abs} on the backgrounds for the antisymmetric tensor
field strengths, one finds that the
following types of (schematically written) couplings are present the
tree-level
superpotential:
\beq
	W_{poss} = \zeta_\C \tr \Phi_{56} + \zeta'_\C \tr \Phi_{34} +
	\theta_a^{a'} Q' \tilde Q
	+ \tilde \theta^a_{a \prime} Q_{a} \tilde Q'
	+ \tr \Phi_{1 2} \theta_a^{a'} \tilde \theta^a_{a \prime}
\eeq
It is clear that nonvanishing $\theta$ and $\tilde \theta$ will give
masses to the quarks as well as $\Phi_{12}$.
In fact, the interacting theory that we are interested in studying
occurs at the origin of parameter space, where all of the parameters
in $W_{poss}$ should be set to zero.  However, we will at present
keep the $\zeta_C$ and $\zeta'_C$ terms and explain why we cannot
turn them on (even as a regulator \cite{abs}) later in the paper.
The complete superpotential is therefore
\beqar	\label {eq:superpot}
	W &=& W_{PP} + W_{PA} + W_{PB} \nono
	&=& \tr \Phi_{12} \comm {\Phi_{34}} {\Phi_{56}} +
	\tr Q \Phi_{56} \tilde Q + \zeta_\C \tr \Phi_{56}
	+ \tr Q' \Phi_{34} \tilde Q'
	+ \zeta'_\C \tr \Phi_{34}.
\eeqar

An important question is whether (\ref{eq:superpot}) is exact.
The symmetries in table 4
show that all the terms which can be constructed from
integer powers
of the fields are already included in  (\ref{eq:superpot}).  One might worry
about
the possibility of negative
or fractional powers.
At large VEVs for $\Phi_{34}$ or $\Phi_{56}$ we should get the instanton
moduli space
as the moduli space of vacua. This suggests that terms with negative or
fractional
powers of the fields are absent.

\section {The Target Space of the Probe}

	In the limit that gravity decouples from the brane theory in
spacetime, the coupling constant of the DLCQ quantum mechanics
becomes infinite.  This corresponds to its infrared (large time) limit,
and the
D0 brane theory
flows to a supersymmetric $\sigma$-model quantum mechanics.  Its
target space is simply the moduli space of flat
directions determined by the usual D-term and F-term equations.
Understanding the geometry and topology of this
space is an important
step in obtaining useful information about the spacetime
theory of tensionless strings using the Matrix probes.

	The target space for the quantum mechanics described in
the previous section possesses very rich structures, not all of which
are relevant to us.  Some parts of the moduli space
describe the probes away from the two types of D4-branes; other parts describe
the probes as instantons in some of the D4-branes.  The only relevant
region for us is the one in which all the probes are stuck at the (1+2)d
intersection.  However, ${\it a~ priori~}$ the D0-brane probes are free to
wander
into other regions of the moduli space.  In this section we analyze
the relevant branch of the moduli space and try to address
this important issue of \emph {decoupling}.
First we review the arguments for decoupling of
the $(2,0)$ theory, which has been studied
extensively in the literature.  The probe theory for that case
has 8 supercharges\footnote
	{The theory of 8 supercharges mentioned in this paper will be
	one that can be obtained from dimensional reduction of a 4d ${\cal
N}$=2
	theory.}.
The branch of interest,
where the D0-branes are probing the interior of the D4-branes, is
the Higgs branch of the moduli space and enjoys strong nonrenormalization
properties that shield its metric from radiative corrections.  Our model
is much more complicated, and is not protected by such powerful
nonrenormalization theorems.  Still,
many similarities to the (2,0) case exist, and
after analyzing the geometry of the moduli space we will be able to
propose and verify a decoupling criterion.

\subsection {A Lesson from the (2,0) Theory}

	The quantum mechanics
which arises in the DLCQ description of
(2,0) theories is a dimensional reduction of $U(N)$ SYM in 4d
with ${\cal N}$=2 supersymmetry.  The theory includes $K$
fundamental hypermultiplets (where $K$ is the number of D4 branes
being probed), and an additional adjoint hypermultiplet.
The F term equations look like

\beq	\label {eq:inst-cuzy}
	\comm {\Phi_{12}} {\Phi_{34}} + \tilde Q Q
	= \zeta_A;
\eeq
\beq	\label {eq:inst-diag}
	\comm {\Phi_{12}} {\Phi} = 0 = \comm {\Phi_{34}} {\Phi};
\eeq
\beq	\label {eq:inst-higgs}
	\Phi \tilde Q = 0 = Q \Phi;
\eeq
$Q$ is a $K\times N$ matrix, $\tilde Q$ is $N \times K$.

The D-term equations for the 4d gauge theory are
\beqar
	\comm {\Phi_{12}} {\Phi_{12}^\dag} +
	\comm {\Phi_{34}} {\Phi_{34}^\dag} +
	\comm {\Phi} {\Phi^\dag} + && \nono
	Q Q^\dagger - \tilde Q^{\dag} \tilde Q 	&=& \zeta_\bR .
\eeqar
where $\zeta_R$ here is the sum of the contribution from the two types
of D4-branes.

To understand the structure of the moduli space of vacua,
we consider solutions to these equations.
We will discuss the 4d gauge theory (i.e. the 3 brane - 7 brane system
instead of the 0 brane - 4 brane system), and then abstract lessons for
the quantum mechanics at the end.
As is well known,
the D-term equation
combined with the $U(N)$ quotient is equivalent to a quotient by $GL(N;C)$
of the holomorphic field variables.  Therefore we only need to study the
solution to the F-term equations quotiented by $GL(N;C)$.  For generic
values of $\Phi$, it is a nondegenerate $N\times N$ complex matrix.
\eqr {inst-higgs} implies that $Q$ and $\tilde Q$ both vanish --- they
become massive.
Also, a generic matrix $\Phi$ has distinct
eigenvalues.
The rest of the F-term equations then imply that we can simultaneously 
diagonalize $\Phi_{12}$, $\Phi_{34}$ and $\Phi$ to use up all of the $GL(N;C)$
except for a $(C^*)^N$ factor, which represents the complexification of the
unbroken
$U(1)^N$ gauge symmetry and the  $S_N$ Weyl group.  This is the
Coulomb phase, the moduli space is that for the adjoint scalar in the
vector multiplet $\Phi$ times
that for the diagonal adjoint hypermultiplet.
The $\Phi$ branch receives quantum corrections to its metric
 from integrating out the
$Q$s and $\tilde Qs$ and develops a logarithmic
``throat,'' while the branch parameterized by the
adjoint hyper retains its classical $R^{4N}$
metric.

	Another branch can be found by letting, say, $\tilde Q$ be generic.
As it turns out, we can also allow, say, $\Phi_{12}$ to be generic.  For
$K \geq N$, genericity of $\tilde Q$ already forces $\Phi$ to be zero
(i.e. it is massive).  For $K < N$, this is insufficient by itself.  However,
\eqr {inst-diag} imposes a condition on $\Phi$.  For a generic value of
$\Phi_{12}$, we can choose a basis in which
$\Phi$ and $\Phi_{12}$ can be simultaneously diagonalized.  In this basis,
it is easy to see that \eqr {inst-higgs} cannot be satisfied for
generic $\tilde Q$ unless $\Phi$ vanishes.  To make this more explicit,
\eqr {inst-diag} and \eqr {inst-higgs} imply that
\beq
	\Phi \exp^{\alpha \Phi_{12}} \tilde Q = 0
\eeq
for arbitrary $\alpha$.  Therefore the $K$ $N$-vectors in $Q$
are in a $\Phi_{12}$
invariant \emph {proper} subspace of $C^N$ unless  $\Phi$ vanishes.
Thus genericity of $Q$ implies the vanishing of $\Phi$.
After setting $\Phi$ to $0$, the rest
of the F-term equations can be satisfied, yielding a moduli space of
complex dimension $2NK$ that is birational to a symmetric product
\cite {witten}.  This is a branch parameterized purely by scalars in the
hypermultiplets, a maximally Higgsed branch. It
is in fact equivalent to the moduli space of $N$ $U(K)$ instantons.

	The above two branches are connected at a region where $Q$, $\tilde
Q$,
and $\Phi$ all vanish.  This is where all the instantons are of zero size but
still attached.  The metric diverges as one travels from the
interior of the Coulomb branch to this point, due to
radiative corrections at one loop from virtual quarks.
On the other hand, the metric from the interior of the
hypermultiplet branch to the origin of $Q$ and $\tilde Q$ and
in varying $\Phi_{12}$ and $\Phi_{34}$ is uncorrected
and remains finite.  There are also mixed branches,
corresponding to non-generic but
nonvanishing values of $\Phi$.  When $\Phi$ has rank $n < N$, for example,
the solutions to the F-term equations give a branch which corresponds  
to $n$ instantons and $N-n$ detached D3-branes.  All these branches are also
separated from each other by ``tubes '' along which the 
metric for $\Phi$ develops similar
divergences.

	The divergence in the metric persists if we dimensionally reduce
this theory to $d < 4$ dimensions.  The divergence of the metric takes the
form $\frac 1 {r^{4-d}}$, while for $d=4$ the tube metric diverged only
logarithmically.  Therefore at $d=2$ and lower the divergence will result
in an infinitely long tube.  This is very important
because in $d \leq 2$, physical states are described by
wavefunctions spread out over the flat directions, and only an infinitely long
tube can decouple wavefunctions in different regions of the moduli space.
In using the D0-branes to probe the D4-branes as in
\cite {abkss,witten,abs}, it
was a necessary sign of consistency that the maximally Higgsed branch
is separated from the mixed and Coulomb branches by such infinite tubes.
Otherwise, the D0-brane
physics could not be used as a definition of an intrinsic theory on the
M5-branes that is believed to decouple from the bulk of spacetime.

	In our theory, we will face a similar test of consistency, but the
reduced supersymmetry allows a moduli space
that is far more intricate.
Because the analysis of decoupling of the utmost importance, and yet
the quantum mechanics for our model is complicated, it is useful to develop
a set of criteria based on the much better understood case of instantons.
Ideally, fluctuations along the flat directions that move the D0 branes
away from the
branch of interest should be massive.  The mass is typically of the
form $g \phi$, where $g$ is the gauge coupling constant and $\phi$
is the VEV that leads to the mass.  In the limit we take, $g \to \infty$ as
$M_{pl} \to \infty$, so the mass actually becomes infinite.
The scalar potential that is responsible for giving masses to
potential moduli takes the form
\beq
	L_{scalar} = \sum_{i} \abs {F_i}^2
\eeq
where $F_i$ are the F-terms.  Therefore massless fluctuations correspond to
the kernel of the matrix
\beq	\label {eq:mass-matrix}
	M_{ij} = \frac {\pa F_i} {\pa \phi_j},
\eeq
where the $\phi$s are chiral multiplets of the theory.
In other words, they are solutions to the linearized F-term equations.

At special subloci on the Higgs branch of interest in our case,
additional fields become massless and the Higgs branch intersects
partial Coulomb branches.  At such points, the decoupling argument
will break down unless the metric on the (partial) Coulomb branch
has an infinite tube due to a loop correction to its metric.
In theories with 8 supercharges, the metric
perturbatively receives at most one-loop corrections.  However, in the
case we study, the reduced supersymmetry allows higher loops to contribute
in general and no general results about perturbatively exact metrics
are known.  Nevertheless,
our model strongly resembles the Matrix description of the
(2,0) theory.  In particular, the superpotential is obtained as
a sum of contributions from different 0-4 subsectors, which each
have 8 supercharges.
At the one loop level,
we can determine whether there is a divergence in the metric
by comparing with the theories of \cite{abkss}.  If there is a divergence at
one-loop, it is unlikely to be removed by higher loops and/or nonperturbative
effects\footnote
{In \cite {ahiss} a scenario like this is conjectured to happen,
but it involves a dynamically generated superpotential.  As explained at
the end of last section, such term is unlikely to appear for the quantum
mechanics we study and
we expect that the divergence persists.}.

	The divergence in the metric generated at one-loop originates from
virtual massive charged particles running in the loop.
Therefore, if along some flat direction the number of
charged massless particles decreases, then we expect a divergent one-loop
renormalization of the metric along that branch.  In other words,
for some solution of
\beq
	M_{ij} \delta_1 \phi_{j} = 0
\comma
\eeq
that corresponds to a charged particle,
there is no solution for $\delta_\epsilon \phi$
to the first order perturbation equation
\beq	\label {eq:cond-for-divergent-metric}
	\frac {\pa M_{i j}} {\pa \phi_k} \delta_2 \phi_k \delta_1 \phi_{j}
	+ M_{ij} \delta_\epsilon \phi = 0,
\eeq
where $\delta_2 \phi$ is the flat direction away from the region of
interest.  
In our case $\delta_1 \phi_j$ is the flat direction along the branch parameterized by $Q, \tilde{Q}$
while $\delta_2 \phi_j$ is the flat 
direction along the branch parameterized by $\Phi_{34}$ and $\Phi_{56}$.
Since by \eqr {mass-matrix}
\[	\frac {\pa M_{i j}} {\pa \phi_k}
	= \frac {\pa M_{i k}} {\pa \phi_j}.	\]
\eqr {cond-for-divergent-metric} is equivalent to
\beq
	\frac {\pa M_{i j}} {\pa \phi_k} \delta_1 \phi_k \delta_2 \phi_{j}
	+ M_{ij} \delta_\epsilon \phi = 0.
\eeq
In other words, this fluctuation $\delta_2 \phi$ also become massive
as one turns on $\delta_1 \phi$.  This is precisely what happens at
the origin of the Higgs branch for the theory of eight supercharges.
Of course, this is only a necessary condition
for the said flat direction to grow an infinitely long tube.  One has to
directly check that the loop graphs including the relevant F-terms lead
to a divergent metric.

\subsection {The Intersecting Fivebrane Quantum Mechanics}

The F-term equations derived from the superpotential \eqr {superpot}
are

\beqar	\label {eq:vortex-cuzy}
	\comm {\Phi_{12}} {\Phi_{34}} + \tilde Q Q
	&=& 0; \\
	\comm {\Phi_{12}} {\Phi_{56}} + \tilde Q^{\prime} Q_{\prime}
	&=& 0;
\eeqar
\beqar	\label {eq:vortex-higgs}
	\Phi_{56} \tilde Q = &0& = Q \Phi_{56}; \\
	\Phi_{34} \tilde Q' = &0& = Q' \Phi_{34}; \\
\eeqar
\beq	\label {eq:vortex-diag}
	\comm {\Phi_{34}} {\Phi_{56}} = 0
\eeq

	Since we want the D0-branes to probe the theory at the intersection,
we want to set $\Phi_{34}$ and $\Phi_{56}$ to $0$.
The reduced F-term equations for the moduli space of interest are
\beqar	\label {eq:vortex-reduced-cuzy}
	\comm {\Phi_{12}} {\Phi_{34}} + \tilde Q Q
	&=& 0; \\
	\comm {\Phi_{12}} {\Phi_{56}} + \tilde Q^{\prime} Q^{\prime}
	&=& 0; 
\eeqar
\beq	\label {eq:vortex-reduced-higgs}
	\Phi_{34} = 0 = \Phi_{56}. 
\eeq
There is no condition on $\Phi_{12}$.

	These equations immediately imply
that for $N>1$, we cannot turn on the complex \FI terms if $N > \max
(K, K')$.
Otherwise, the RHS of \eqr {vortex-cuzy} would be replaced by a nonvanishing
multiple of the identity.  Then, the RHS would
have greater rank than the maximum possible for the
LHS and there is no solution.  Intuitively, a complex
\FI term forces the instantons to spread out on the D4-brane
they are associated with in all directions, which directly conflicts with
confining the D0-branes in the intersection.
This is in stark contrast to the $(2,0)$
model analyzed in \cite {abs}.  However, the real \FI terms are
still compatible
with our solution.  We will use
them to resolve certain singularities, and analyze the
compact
cohomology of the (partially desingularized) resultant space, as in
\cite{abs}.
We will return to this point later on.

	The solution space to the matrix equation $\tilde Q Q = 0$
has a series of branches, labeled by an
integer $\lambda = rank (Q)$ ranging
 from $\max (0, K - N)$ to $\min(N,K)$ inclusive, with dimensions
\beq
	N K + \lambda K  - \lambda^2 . 
\eeq
The total space is quite complicated, with these different branches
emanating from positive codimensional submanifolds.  It is unlikely
that the branches are separated by an infinite distance.
A similar structure exists in the instanton moduli spaces which arise
in the DLCQ definition of the (2,0)
theories, and there all of these branches must be included.
We study the details of this space in the following subsections for
different values of $K$, $K'$, and $N$.

    To find the total dimension of the moduli space
after quotienting
by $GL(N;C)$, we diagonalize $\Phi_{12}$.
The eigenvalues of $\Phi_{12}$ then contribute $N$ to the dimension of
the moduli space.
Adding the contributions of the particular branch we choose
for the $Q$ and $\tilde Q$ VEVs and the
$Q'$ and $\tilde Q'$ VEVs, and finally taking into account
the quotient by the remaining
$C^*$s, the dimension of the branch characterized by
$(r = \Rank (Q), r' = \Rank (Q'))$ is
\beq
	N + (NK + rK- r^2) + (NK' + r' K' - r'^2) - N
	= N (K+K') + r(K-r) + r' (K-r').
\eeq

	As we shall discuss in the next section, for
a Matrix quantum mechanical model to have
the usual DLCQ probe interpretation, the cohomology of its moduli
space should exhibit the properties of a symmetric product.
The simplest way this can happen is if
the moduli space itself
is a symmetric product.
This however turns out not to be true in our case and the moduli space is not a symmetric product.
We shall comment on this as we analyze
the moduli space for the appropriate case and return for a more
elaborate discussion in the next section.

To justify the constraint \eqr {vortex-reduced-higgs} we also need to
analyze the solutions to the linearized F-term equations
with $\Phi_{34}$ and $\Phi_{56}$
set to zero:
\beqar	\label {eq:linear-cuzy}
	\comm {\Phi_{12}} {\delta \Phi_{34}} + \delta \tilde Q Q
	+ \tilde Q \delta Q &=& 0; \\
	\comm {\Phi_{12}} {\delta \Phi_{56}} + \delta \tilde Q' Q'
	+ \tilde Q' \delta Q' &=& 0;
\eeqar
\beqar	\label {eq:linear-higgs}
	\delta \Phi_{56} \tilde Q = &0& = Q \delta \Phi_{56}; \\
	\delta \Phi_{34} \tilde Q' = &0& = Q' \delta \Phi_{34}.
\eeqar

\subsubsection {$K = K' = 1$}

	Let us first consider the simplest case, $K = K' = 1$.
The equation for $\tilde Q_i Q^j$
has two branch of solutions: one in which $Q = 0$ but $\tilde Q$ is 
arbitrary,
and the converse.  The same is true for $Q'$ and $\tilde Q'$.
$\Phi_{12}$ is arbitrary.  We must now take into account the
$\GLNC$ quotient.  Generically, we can diagonalize $\Phi_{12}$.  This
gauge-fixes $\GLNC$ up to a residual ${C^*}^N \Join S_N$, where $\Join$
denotes
a semi-direct product and $S_N$ is the N-th order permutation group.

	To gauge-fix the residual symmetry, we consider the four branches 
of solutions to \eqr {vortex-reduced-cuzy} as in \tabr {simple-four-branches}.
\btab {simple-four-branches}
{Four branches of solutions to \eqr {vortex-reduced-cuzy} for $K = K' = 1$}
{ l  c  c  c  c }
Branch & & $Q = 0 = \tilde Q'$ & $\tilde Q = 0 = Q'$
	$Q = 0 = Q'$ & $\tilde Q = 0 = \tilde Q'$ \\ \hline
After Quotient & $(\C)^N$ & $(\C)^N$ & origin & origin \\ \hline
\etab
Each $C^*$ factor of the residual symmetry group acts on a pair of primed
and unprimed quarks.  If one is tilded and the
other untilded,
then $C^*$ acts instead as, for example,
\[	(Q, \tilde Q') \to (\lambda Q, \frac 1 {\lambda} Q')	\]
and the quotient is still noncompact and simply $C^1$, parameterized
by $Q\tilde Q'$.
If they are of the same type, (both tilded or both untilded)
then $C^*$ acts as, for example,
\[	(Q, Q') \to (\lambda Q, \lambda Q').	\]

Naively, the result of this quotient is a $\CP^1$ with the pair as
the homogeneous coordinates.  However, there is a subtlety.
In \cite {luty-taylor}, where the D-flatness condition with vanishing
real \FI term is considered, it was found that
the $C^*$ quotients must be taken in a generalized sense so that,
in the present context, the whole $\CP^1$ is collapsed to and identified
with the point at the origin ($Q = \tilde Q = Q' = \tilde Q' = 0)$.
Finally we take into account of the $S_N$ action.  This just replaces the
direct products in \tabr {simple-four-branches} by symmetric products.
Therefore the
moduli space looks like \figr {only-cones}.
\putfig {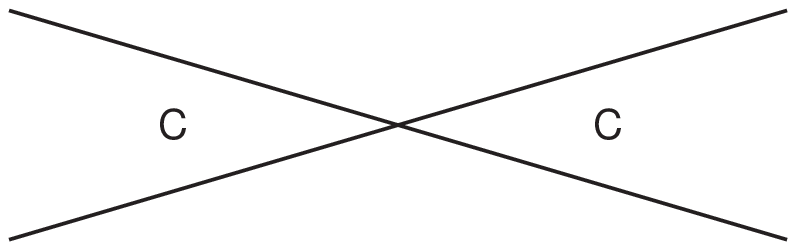} {\FigSize}
	{Moduli space for $K=K'=1$, $\zeta = 0$}
It consists of
two cones, each a symmetric product of $N$ $\C$'s.
If one turns on $\zeta_R$, the situation becomes subtler still.
Now one of the two branches on the right in \tabr {simple-four-branches}
is lifted completely while the other become a compact space, replacing the
origin.  For $N = 1$, the compact space is a $\CP^1$ and the moduli
space looks like \figr {cones-and-sphere}
The size of the sphere is
controlled by $\zeta_R$.  Note that the sphere still touches the two
$C$s at a point.
\putfig {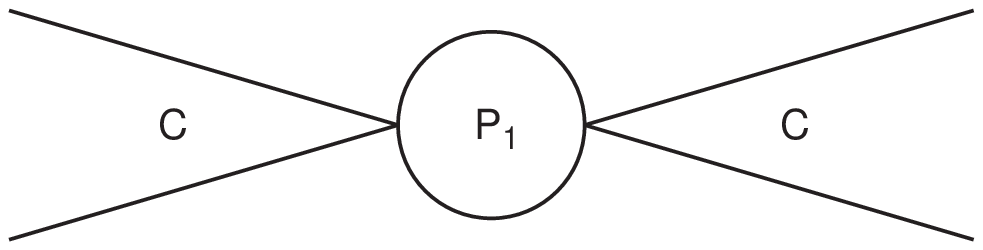} {\FigSize}
	{Moduli space for $K = K' = 1 = N$, $\zeta \neq 0$}
For higher $N$, the sphere is replaced by some other complex N 
manifold.

	Now we shall justify imposing the constraint
\eqr {vortex-reduced-higgs} by demonstrating decoupling in the manner
outlined earlier.
The same argument works for all four branches and for both $\Phi_{34}$
and $\Phi_{56}$, so for definiteness we take $Q = 0 = Q'$, $\tilde Q$ and
$\tilde Q'$ arbitrary, and concentrate on $\Phi_{34}$.
The relevant linearized F-term equation is
\beq	\label {eq:example-one-one-higgs}
	\Phi_{34} \delta \tilde Q' = 0. 
\eeq
Let us diagonalize $\Phi_{34}$ and
 look at \eqr {linear-cuzy}, which specializes
to
\beq	\label {eq:example-one-one-cuzy}
	\comm {\Phi_{12}} {\delta \Phi_{34}}
	+ \tilde Q \delta Q = 0. 
\eeq
The commutator's diagonal elements are all zero.  Generically $\tilde Q$
is a column N-vector with all nonvanishing entries.  The two together imply
that
\[	\delta Q = 0,	\]
which means $\delta \Phi_{34}$ must commute with $\Phi_{12}$, an arbitrary
$N\times N$ matrix.  As shown earlier, this together with
\eqr {example-one-one-higgs} imposes a condition on $\Phi_{12}$ and
$\tilde Q'$ that is not satisfied generically.  For nongeneric
$\Phi_{12}$ and $\tilde Q$ that does support a solution to
\eqr {example-one-one-higgs} and \eqr {example-one-one-cuzy},
the same argument as in the case of the (2,0) theory
suggests that there is an infinite tube along the flat direction of
$\Phi_{34}$ and completes the decoupling argument.

\subsubsection {General Case --- Moduli Space}

	The analysis of the moduli space for higher $K$ and $K'$
proceeds in a similar vein, but becomes very complicated.  First of all,
the solution space to $\tilde Q Q = 0$ now consists of more than two
branches.  The additional possibility is for $Q$ and $\tilde Q$
nonvanishing.  As one would expect from 4d ${\cal N}$=1 theories, the
structure
changes drastically for $N$ greater or less than $K$ and/or $K'$.
Perhaps more surprisingly, the discussion of decoupling also depends on
$N$.  There may be a spacetime connection between the two.
We take this as a suggestion that for the theories we
study here the probe reflects the chiral primary spectrum of the
spacetime theory only in the large N limit, contrary to the
case studied in \cite {abs}.

\paragraph {$K \geq K' \geq N$}

Once $K$ and/or $K'$ is greater than $1$, the solution to
\eqr {vortex-reduced-cuzy} becomes complicated, as explained
in the paragraph following that equation.  When $N < K \leq K'$,
the space of solutions does not 
contain any symmetric product.  To see this, consider
for illustration the equation $\tilde Q Q = 0$ and think of
$Q$ and $\tilde Q$ each as $N$ $K$-vectors grouped together.
Because $N < K$, both $Q$ and $\tilde Q$ can be simultaneously nonvanishing
in any branch.  Generically in each branch, the space can
be thought of as the solution for $\tilde Q$ fibered
over a matrix $Q$ of certain rank or the converse.  The fiber
is nontrivial and become singular (enlarged) at certain submanifolds of
the base.  It is this fibration structure that prevents the total
space from becoming a symmetric product after taking the $S_N$ quotient.
For $K < N$, this fiber is nontrivial in every branch.

	On the other hand, the decoupling of $\Phi_{34}$
and $\Phi_{56}$ is straightforward.  The genericity of $Q$
and $\tilde Q$, for instance,
in each branch is sufficient to force both
$\Phi_{56}$ and $\delta \Phi_{56}$ to
vanish through \eqr {vortex-higgs}, because together they form $K \geq N$
linearly independent $N$-vectors.  The arguments given earlier show
that at nongeneric values of $Q$ and $\tilde Q$, the flat direction
along $\Phi_{56}$ is an infinitely long tube.  Similar arguments apply
to $\Phi_{34}$.  However, despite the decoupling, the spacetime
interpretation of such cases is not clear, because of the lack of
a symmetric product structure.  It is possible that for this model we
should only expect the usual spacetime interpretation to hold at
sufficiently large N, while for smaller N some peculiarity of
the DLCQ becomes significant and obscures the spacetime physics.

\paragraph {$N > K \geq K'$}

	For $N \geq K$, there are branches
in the solution space to $\tilde Q Q = 0$ where either $Q$ or $\tilde Q$
is constrained to be zero.  The case $K = 1$ treated earlier is a special
case of this type.  These
are the branches in which the fibration structure discussed above
become trivial.  Therefore, as discussed earlier,
the whole moduli space will have components
that are symmetric products after taking
care of the quotient.
Because this is the appropriate structure for a spacetime
interpretation, we shall concentrate on these components.
The whole analysis parallels that for $K = K' = 1$ and we shall be brief.

	These well behaved branches again decompose into four components.
They and their quotient by $(C^*)^N$
have a similar classification to the one we found in table 6, as shown in 
\tabr {complex-four-branches}.
\btab {complex-four-branches}
{Four branches of solutions to \eqr {vortex-reduced-cuzy} for $K = K' = 1$}
{ l  c  c  c  c }
Branch & $Q = 0 = \tilde Q'$ & $\tilde Q = 0 = Q'$
& $Q = 0 = Q'$ & $\tilde Q = 0 = \tilde Q'$  \\ \hline
After Quotient & $(\C^{K+K'-1})^N$ & $(\C^{K+K'-1})^N$
	& origin & origin  \\ \hline
\etab
After $\zeta_R$ is turned on, a compact space emerges and replaces the
origin in similar fashion to what happens for $K = K' = 1$.

	The decoupling analysis also becomes more involved.
\eqr {vortex-higgs} by itself
can force neither $\Phi_{56}$ nor $\Phi_{34}$
to be zero even generically, because the quarks do not have sufficient rank.
As in the
$K = K' = 1$ case, one has to use \eqr {linear-cuzy}.   Let us concentrate
on $\Phi_{34}$ and the branch in which $Q = 0$.  \eqr {linear-cuzy}
reduces to
\beq	\label {eq:large-N-cuzy}
	\comm {\Phi_{12}} {\delta \Phi_{34}}
	= - \tilde Q \delta Q.
\eeq
This
time $\tilde Q \delta Q$ does not necessarily vanish.
However, because $K < N$, the RHS has only rank K and is not generic
and is constrained by the value of $\tilde Q$.  The LHS, however,
depends only on $\Phi_{12}$, $Q'$, and $\tilde Q'$ -- the latter two
through
\beq	\label {eq:large-N-higgs}
	\delta \Phi_{34} \tilde Q' = 0 = Q' \delta \Phi_{34}.
\eeq
 From \eqr {vortex-reduced-cuzy} and
\eqr {vortex-reduced-higgs}, we see there is no correlation between them.
Therefore generically there is no solution to
\eqr {large-N-cuzy}.  The same applies to
$\Phi_{56}$.

\paragraph {$K > N > K'$}

	For values of $N$ in this range, the moduli space is a mixture
of the two types we analyzed above.
The solution space for $Q'$ and $\tilde Q'$ has components that
are N-th order symmetric products, while that for $Q$ and $\tilde Q$ does
not.
Since $S_N$ acts simultaneously on primed and unprimed quarks, the total
space does not have the structure of a symmetric product.

	The analysis of decoupling for these cases is
even more intricate.  While
\eqr {linear-higgs} forces $\delta \Phi_{56}$ to vanish,
it only forces $\delta \Phi_{34}$ to have rank
no more than $N - K'$.  Does \eqr {linear-cuzy} impose any constraint?
At least in some branches of moduli space the answer is no.
Consider the one in which the rank of $Q$ is
$N$.   Then for arbitrary $\delta \Phi_{34}$,
there is a solution to \eqr {linear-cuzy} for $\delta \tilde Q$
because $Q$ is invertible in an appropriate sense.  $K \geq N$ is
crucial here.  For other
branches the situation is more complicated and it appears at least for some
of them \eqr {linear-cuzy} imposes a very weak constraint on
$\delta \Phi_{34}$.  Nonetheless for these values of $N$, the
problems in at least
one part of the Higgs branch strongly suggest a difficulty with
decoupling.


\section{Moduli space and chiral primary operators}

In previous sections, we have developed a
quantum mechanical matrix 
description of the four dimensional ${\cal N}=2$ superconformal
theory on the intersection of $K$ and $K'$ fivebranes.  In 
this section, we use this quantum mechanics to
compute the dimensions of some of the chiral primary operators
in the CFT with tensionless strings.

The target space of the quantum mechanics describing the
theory on the intersection of the branes has singularities of various codimensions.
 We resolve them  by using the  FI parameters.
We have two sets of FI parameters 
$(\zeta_{R},\zeta_{C})$ and $(\zeta_{R}',\zeta_{C}')$.
As in \cite{abs}
the space-time interpretation of the FI parameters involves turning on constant self dual
3-form field strength $H ,H' $ on the two sets of fivebranes.
We have 
$(\zeta_{R},\zeta_{C})  \sim H_{+ij}, i,j=1,2,3,4$ and 
$(\zeta_{R}',\zeta_{C}') \sim H_{+ij}', i,j=1,2,5,6$.

Turning on the real FI parameters corresponds to
a resolution of singularities. Turning on the complex FI parameters corresponds to
a deformation.
We argued in the previous section that in order to have a consistent
space time interpretation of the theory  we have to set $\zeta_C = \zeta_C' =0$.
We will further elaborate on this point later.
The superconformal field theory that we study arises when the FI parameters vanish.
As in \cite{abs} we consider the large time behaviour of the quantum mechanics 
where the finite FI parameter can be absorbed by a wave function renormalization of the operators.
This provides a procedure to relate the quantum mechanical wave functions at finite FI parameters
to the states of the SCFT.

We will study the
chiral primary operators of the
four dimensional theory that  correspond to  the compact cohomology of the resolved moduli space
 of the 
quantum mechanics localized at the origin \cite{abs}.
The localization at the origin is a consequence of the general fact that a
 quantum mechanical 
state which corresponds to a  primary
operator is  localized at the origin of the moduli space \cite{abs}, and this will play an important role
in the analysis.
This state may be viewed as  a particular representative of the compact cohomology which is obtained 
by scaling of a compact cohomology representative which is not concentrated at the origin.

The space-time dimension $D$ of  a chiral primary operator corresponding to a form $O$ with compact
support of degree 
$deg(O)$ 
is 
\beq 
D = deg(O)  - \frac{1}{2} dim_{R} ( \cM^{res}_{N;K,K'} )
\label{D}
\stop
\eeq 
$\cM^{res}_{N;K,K'}$ is the resolved moduli space. 
\subsection{ The case $K=1, K'=1$}

Consider  the system with $(K=1,K'=1)$.
The moduli space $\cM_{N;1,1}$ has complex dimension $2N$.
It has the structure of a product of  $C^N$ with the 
$N$ complex  dimensional space constructed from 
two copies of $C^N/S_N$ intersecting at the origin.

We will first analyze the momentum one case $N=1$. 
The moduli space  $\cM_{1;1,1}= C \times \cM_{1;1,1}^0$ has complex dimension two.
$\cM_{1;1,1}^0$ is parameterized by the complex coordinates
 $A=Q\tQ', B=Q'\tQ'$ and has the structure of two complex planes intersecting at the origin as
in figure 1.
Therefore the origin is a singular point.
Note for comparison that in the $(2,0)$ field theory on $k$ parallel fivebranes studied in \cite{abs}
 the $k=1,N=1$ 
moduli space is $R^4$ which does not have a singularity. 
Turning on the real FI parameter $\zeta_{R}$ changes $\cM_{1;1,1}^0$ to the space of figure 2.
This  does not resolve the singularity at the origin
 but rather
shifts its location to two points. In order to resolve the singularity at the origin  we
have to deform it using a complex FI parameter $\zeta_{C}$.
However the singularity at the origin  is part of the physics that one D0 brane probes,
and a deformation of this singularity will change the physics of the system.

Consider the 
compact cohomology of 
$\cM_{1;1,1}^0$ scaled to be localized at the origin. 
Obviously we have  the top 2-form localized at the origin.
The question is whether we have something else. 
Although there is a 0-form on the $CP^1$ connecting the two complex planes in figure 2, it 
cannot be extended to a 0-form on the 
whole $\cM_{1;1,1}^0$. The reason is that
its extension would have to be a constant and the moduli space is noncompact.  
We could also 
try to consider the top 2-form on each of the complex planes or on the $CP^1$ separately.
However such 
a compact cohomology element  has to vanish on the points of intersection of the complex planes
and the $CP^1$, which means that is vanishes as $\zeta_R \rightarrow 0$. Therefore 
such candidate forms cannot represent
wave functions localized at the origin.

To summarize,  the compact cohomology localized at the origin of
$\cM_{1;1,1}$ consists of only the top 4-form which is the wedge product of the
top form of $C$ and the top form of $\cM_{1;1,1}^0$.
The top 4-form localized at the origin corresponds to a chiral primary operator $O$ of the four dimensional theory.
Using (\ref{D}) the dimension of the 
operator $O$ is two.
The fact that we have a dimension two operator implies that the space time
theory  is a non-trivial 
${\cal N}=2$ SCFT.
The operator $O$ is part of the ${\cal N}=2$ chiral ring at the point where the string
becomes tensionless. 
This can be contrasted with the case of one fivebrane. 
In this case the analysis
of the compact cohomology 
in \cite{abs} shows that there is one operator of dimension 2
in six dimensions and the theory is free. 
Our case resembles more the case of $k=2$ fivebranes
in \cite{abs}. In that case there is also a tensionless string and the theory is not free
in the IR.

Consider next the momentum two case $N=2$.
The moduli space  $\cM_{2;1,1}= C^2 \times \cM_{2;1,1}^0$ has complex dimension four.
$\cM_{2;1,1}^0$ consists of two copies of $C^2/S_2$ 
intersecting at the origin.
The analysis is similar to the $N=1$ case. Turning on $\zeta_R \neq 0$ connects the two 
 $C^2/S_2$ spaces by a compact space.
Primary operators correspond to quantum mechanical states localized at the origin.
In order to describe chiral primary operators  we will consider compact cohomologies
on each of the components of $\cM_{2;1,1}^0$ and scale them to the origin. 
We cannot consider compact cohomologies on each of the components separately since we will have
to require that they vanish on the intersections which at $\zeta_R \rightarrow 0$ implies that they
vanish at the origin.
Therefore we get a compact cohomology class 
corresponding to the resolution of the orbifold singularity of the two $C^2/S_2$.
It gives  a 6-form on $\cM_{2;1,1}$ which we scale to be localized at the origin.
We also have the top  8-form which we can localize  at the origin.
Thus we get using  (\ref{D}) two operators of dimensions  
two and four.
The interpretation of the result is clear: 
at  $N=2$ we expect to see
$O(1/R)^2$ and $O(2/R)$ which are of dimensions four and two respectively.

The generalization to arbitrary momentum $N$ is straightforward.
The moduli space $\cM_{N;1,1} = C^N \times \cM_{N;1,1}^0$ where  $\cM_{N;1,1}^0$
 consists of two copies of $C^N/S_N$ 
intersecting at the origin.
Turning on $\zeta_R \neq 0$ connects the two 
 $C^N/S_N$ spaces by a compact space.
In order to describe chiral primary operators  we will consider compact cohomologies
on each of the components of $\cM_{N;1,1}^0$ and scale them to the origin. 
Again, we cannot consider compact cohomologies on each of the components separately since we will have
to require that they vanish on the intersections which at $\zeta_R \rightarrow 0$ implies that they
vanish at the origin.
Thus, we find that the number of chiral primary operators of dimension $2k$ is 
\beq
b_{2k} = p_k(N),~~~~~~~~k=1,...,N
\label{coh}
\comma
\eeq
where $p_k(N)$ is the number of partitions of $N$ to $k$ parts.
Note that a consistency check 
on our analysis is the fact that we do not get operators with negative
dimensions.
The result is compatible with the fact that at momentum $N$ we expect
to see operators that   are  products of $O$  at momenta $p_{-}= k_i/R$, namely
$O(k_1/R) \cdots O(k_r/R)$ where $\sum k_i = N$.

If we deform  $\cM_{N;1,1}$ using the complex FI parameters 
then at $N=1$ we get $AB= \zeta_C$, and   compact cohomology of the deformed
$\cM_{1;1,1}^{def}$  localized at the origin consists of a  3-form and a 4-form.
Using (\ref{D}) the dimension of the  corresponding chiral primary
operators $\Psi, O$ are one and two respectively.
However for higher $N$ we lose the symmetric product structure and the 
the analysis of the compact cohomology localized at the origin  is 
not compatible with a space-time interpretation.  This is in agreement with
the fact that turning on $\zeta_C$ forces the probes to spread out
off of the intersection,
and should therefore not correspond to a spacetime deformation of the
3+1 dimensional theory.  

So, we see that requiring a family of quantum 
mechanical systems to be the DLCQ description of a space time theory is very stringent.
It requires that 
the appropriate compact 
cohomologies 
scaled to the origin have the dimensions of classes  
arising from the resolution of orbifold singularities of a symmetric product target space.
In our case, where the moduli space had two components intersecting
at the origin, this implied that we should not resolve the singularity at the origin.

\subsection{The general $K,K'$ case}

Consider now the general system with arbitrary numbers of fivebranes
$(K,K')$.
The structure of the  moduli space $\cM_{N;K,K'}$ has been analyzed in 
section 3. It has several branches of different dimensions.
One of the branches has complex dimension $N(K+K')$ 
and its structure is a direct generalization
of that in the case $K=K'=1$. 
It has the structure of a product 
of  $C^N$ with the the $N(K+K'-1)$ complex  dimensional space constructed from 
two copies of $C^{N(K+K'-1)}/S_N$  intersecting at the
origin.
Again, the real FI parameter  resolves the singularities of the symmetric products and connects 
the two  $C^{N(K+K'-1)}/S_N$ spaces by a compact space .
The other branches have various dimensions that depend on an integer parameter $\lambda$ 
and do not have the form of a symmetric
product.

As before we want to consider quantum mechanical states that are localized at the origin. In the case $K=K'=1$
we considered compact cohomologies scaled to the origin.
This led us to consider  only those compact cohomologies that arise from the
resolution of the symmetric product singularities and the top form.
All other possibilities of compact cohomologies (localized on one of the two resolved $C^{N}/S_N$s 
or on the compact space arising in the $\zeta_R \neq 0$ case) vanish upon scaling to the origin.
Applying the same strategy here we are led to consider only the compact cohomology
classes which arise 
from the resolution of symmetric product singularities on 
the branch of complex dimension $N(K+K'-1)$.

Consider first the momentum one case $N=1$. 
The moduli space  $\cM_{1,K,K'}= C \times \cM_{1;K,K'}^0$ has complex dimension $K+K'$, where 
$\cM_{1;K,K'}^0$ has the structure of 
two copies of $C^{K+K'-1}$  intersecting at the
origin.
 Turning on $\zeta_R \neq 0$ connects the two 
 $C^{K+K'-1}$ spaces by a compact space.
Again, in order to describe chiral primary operators  we will consider compact cohomologies
on each of the components of $\cM_{1;K,K'}^0$ and scale them to the origin. 
We cannot consider compact cohomologies on each of the components separately since we will have
to require that they vanish on the intersections which at $\zeta_R \rightarrow 0$ implies that they
vanish at the origin.
Therefore we get  
only the $2(K+K')$ top form of 
$\cM_{1;K,K'}$
localized at the 
origin. It  corresponds to a chiral primary operator $O$ of the four dimensional theory of
 dimension $K+K'$.

The generalization to arbitrary momentum $N$ is straightforward.
We look for compact cohomologies localized at the origin.
Only those that arise 
from the resolution of the symmetric product singularities are relevant.
We get that the number of chiral primary operators of dimension $k(K+K')$ is  
$p_k(N),k=1,...,N$.
As guaranteed by the procedure 
the  result is compatible with the fact that at momentum $N$ we expect
to see operators that   are  products of $O$  at momenta $p_{-}= k_i/R$, namely
$O(k_1/R) \cdots O(k_r/R)$ where $\sum k_i = N$.

\section{General moduli space}

An interesting question that naturally 
arises is the following: Given supersymmetric quantum
mechanics on 
a moduli space $\cM_{N;\vec{k}}$ where $\vec{k}$ stands for a set of parameters
$\vk =\{k_1,...k_n\}$ 
what are the necessary and sufficient conditions to have a space time 
interpretation of the model in the DLCQ sense.
In the following 
we will discuss necessary conditions on the  
cohomology with compact support
(and localized 
at the origin). 

In general we tend to expect  that $\cM_{N;\vec{k}}$ is birational to the symmetric
product of $\cM_{1;\vec{k}}$
\beq
\cM_{N;\vec{k}} \simeq Sym^N \cM_{1;\vec{k}}
\label{bir}
\comma
\eeq
namely they are equivalent up to the singularities. The physical reason behind this
is the fact that $N$ D0 branes probing the space-time theory see the $N$-th product of   
the moduli space seen by one D0 brane, up to permutation.
However, in the examples that we studied in previous sections we saw that this is not necessarily
the case. The physics 
of N D0 probes can sometimes involves couplings that do not vanish even if we separate 
them. 

The real requirement 
involves the compact cohomology localized at the origin, which corresponds 
to chiral primary operators.
In the case that  $M_{N;\vec{k}}$ satisfies (\ref{bir}) these correspond to  
the  compact cohomology of the resolved space.
Let us proceed with this case and discuss the more subtle case later.
Consider  $\cM_{1;\vec{k}}$ and 
let $d$ be the complex dimension of $M_{1;\vec{k}}$.
Since the compact cohomologies of $M_{N;\vec{k}}$
correspond to chiral primary operators in the space time
theory
with dimensions given by (\ref{D}), and these dimensions must be non-negative, 
we have as a second requirement that the compact cohomologies satisfy
\beqar
{\rm dim}  H^p(M_{1;\vec{k}}) &=&0~~~~~~p=0,...,d-1 \non\\
{\rm dim}  H^p(M_{1;\vec{k}}) &=&b_p~~~~~~p=d,...,2d
\comma
\label{coho}
\eeqar
where $b_p$ can be different than zero.

Consider now  general $N$. 
We expect that the chiral primary operators that we see at momentum $N$,
 which we get from the compact cohomology
of $\cM_{N;\vk}$, are just products of the operators
$O_i$ at momenta $p_{-}= k_i/R$, namely
$O_i(k_1/R) \cdots O_r(k_r/R)$ where $\sum k_i = N$.
Using this  fact and (\ref{coho}) we can derive  
the generating formula for the dimensions of the compact
cohomologies 
\beq
\prod_{l=1}^{\infty} 
\prod_{i=d}^{2d} \left(1-(-1)^i t^{i+d(l-1)}q^l \right)^{-(-1)^i b_i}
= 
\sum_{p,N} {\rm dim} H^p(\cM_{N;\vk}) t^p q^N
\stop
\label{general}
\eeq
This is the third requirement.
Note that due to (\ref{coho}) we have 
${\rm dim} H^p(\cM_{N;\vk}) = 0, p < Nd$.

In (\ref{general}) we considered the general case with both odd and even
cohomologies. The moduli space  that we studied in previous sections (as well
as the moduli space of $k$ $U(N)$ instantons) has only even cohomology classes.
Note that when the cohomology is odd (even) we have the corresponding 
term in (\ref{general})
in the numerator (denominator). The interpretation of this in the quantum mechanics
is that while the chiral primary operators corresponding to even cohomologies have
bose statistics, those that correspond to odd cohomologies have odd statistics.
Thus, for instance $O(k/R)O(k/R)=0$ if $O$ corresponds to an odd-dimensional 
cohomology class.

When the moduli space 
does not satisfy (\ref{bir}) the above discussion holds provided in (\ref{coho})
and (\ref{general}) we replace 
${\rm dim} H^p(\cM_{N;\vk})$ by the compact cohomology localized 
at the origin, as discussed 
in previous sections.

\section{Discussion}

Our results suggest that there is a decoupled theory living on the 3+1 dimensional
intersection of groups of $K$ and $K'$ M5 branes.  The part of the chiral ring we 
are able to identify with the cohomology of the Higgs branch seems to be generated by
a single chiral operator, of dimension $K + K'$.  This might seem surprising in view
of the fact that the (2,0) theory living on $K$ coincident M5 branes has a spectrum of 
independent chiral operators that grows with $K$ \cite{abs}.  
However, in that case the field theory has a moduli space
$R^{5k}/S_k$ (along which one separates the $K$ M5 branes) 
and the states in the quantum mechanics have a natural interpretation as
functions on this moduli space.   In contrast, in our case   
separating one of the $K$ (or $K'$) M5 branes from the rest is related to a 
field theory mode supported on the full $K$ ($K'$) M5 brane theory, and does
not correspond to an excitation which is localized on the intersection.  Presumably,
only motion of the center of mass of the $K$ M5 branes away from the $K'$ M5 branes
is related to a state in the decoupled quantum mechanics.  This is the state created
by an operator of dimension $K + K'$.

In the DLCQ description of the e.g. (2,0) field theory \cite{abkss}, decoupling was
obvious for every momentum sector, even for small values of $N$.  In contrast,
here we find that only for $N$ large enough (compared to $K$ and $K'$) can one make
a decoupling argument.  Similarly, in the (2,0) case the moduli space
$\CM_{N,k}$ was (birational to) a symmetric product $(\CM_{1,k})^N/S_N$.  
In our case the moduli space for generic $K,K'$ is not a symmetric
product.  However, at large $N$ the $\it cohomology$ of this space, which is
related to chiral operators in spacetime, does look like that of a symmetric product.
Since the DLCQ is really only ``supposed'' to be related to the higher-dimensional field
theory at large $N$, this is not unacceptable.  

Recently, problems which arise in the DLCQ of 4d field theories (due to strongly
coupled zero modes) were discussed in \cite{hellpol}.  Our description, like
analogous descriptions of other field theories 
derived using D0 brane probes in string theory, does not
obviously suffer from these problems (in much the same way that the D0 brane
quantum mechanics of M(atrix) theory does not suffer from the problems a direct  
DLCQ of 11d supergravity would).  
Nevertheless, the difficulties we encounter at low values of $N$ may be somehow
related to the issues discussed in \cite{hellpol}.  

Finally, we should note that recently it has been proposed (and to some extent
verified) that one may use supergravity as a master field to solve certain
conformal field theories which arise on branes, if the supergravity solution is
known \cite{malda}.   However, for the case of intersecting branes 
(or branes ending on other branes), the appropriate (localized) supergravity
solutions are not yet available.  Therefore, the DLCQ approach is still the 
primary tool we currently have for investigating these theories.

\section*{Acknowledgments}
We would like to thank
O. Aharony, T. Banks, M. Berkooz and E. Silverstein for discussions.
S.K. would like to acknowledge the hospitality of the Institute
for Theoretical Physics in Santa Barbara while this work was being
completed.
This work was supported in part by
NSF grant PHY-951497 and DOE grant DE-AC03-76SF00098.
S.K. was also supported by a DOE Outstanding Junior
Investigator Award and NSF grant PHY-9407194 through the
Institute for Theoretical Physics.
Z.Y. is supported in part by a Graduate Research Fellowship of
the U.S. Department of Education.

\end{document}